# Growth Rate of Staphylococcus Aureus in weak magnetic field

Samina Masood

Department of Physical and Applied Sciences,

University of Houston Clear Lake, Houston, TX 77058

E.mail: masood@uhcl.edu

**Abstract:**

A comparative study of the growth of *S.aureus* in different types of weak magnetic field shows that the growth of the bacteria is initially suppressed after inoculation in a new medium and then increased in the same pattern with or without magnetic field. However, the rate of growth and death of bacteria depend on the variation in the magnetic field for comparable strength. Magnetic field effect on the ionic motion plays a key role in the growth rate. It is shown that the concentration of bacteria reduces initially and then it starts increasing after some time. Multiplication and death of bacterial cells is related to its interaction with the nutrients which is managed by the magnetic field direction, strength and variation at room temperature.

**Introduction:**

*S.aureus* is a typical gram positive bacteria with a disk-like (coccus) shape. Diameter of the bacterial disk is around 1micrometer and is known to make a grape-shape cluster. *S.aureus* can be distinguished from other species of staphylococcus from its interaction with Coagulase [1]. *S.aureus* may or may not be Coagulase-positive [1, 2]. However, it causes several types of infections [3, 4]. These infection may vary from minor skin infections to life-threatening infections causing blood clotting. Moreover, thick cell walls of gram positive *S.aureus* may also develop resistance to antibiotics.

Magnetic field affect the mobility by interacting with the weak negative charge on bacteria. Charged nutrient ions mobility will affect the growth rates and the bacterial concentration for the same composition of nutrient broth. Preliminary study shows that the interaction of antibiotics with bacteria is affected by the weak magnetic field [5] as well.

The positive and negative charge on ions in nutrients and small negative charge on bacteria respond to the Eddy currents produced by the magnetic field. Motion of these charges by the magnetic field varies with mass and charge of ions. Ionic motion is more affected by the magnetic field whereas Eddy currents may not be strong enough to affect the mobility of bacterial cells as much as they transport nutrients due to smaller mass and larger charge. Eddy current will therefore control the relative motion between cells and ions and have significant impact on interaction and absorption of nutrients with cells. The interaction of bacterial cell wall and the absorbance of ions is also affected by the magnetic field.

## Methods and Materials

The static magnetic field was created with the collection of 10 small bar magnets arranged together to create a magnetic sheet [6] and particular locations were identified to put the test tubes straight on those locations. We identify this field as a static field in this paper. The random magnetic field was generated by winding the magnetic strip in the form of a disk. This disk has a magnetic field anywhere between zero and ± 50 Gauss. We try to select fixed points on the disk but the variation of field is unavoidable even between the neighboring points. We designed a slowly damping magnetic field, generated by a 6V battery connecting to a copper coil created around a plastic spool which can individually hold a test tube straight in it. Just to keep everything straight, coils and batteries were set in beakers and filled with Styrofoam pieces to keep them straight throughout. All the arrangements of the magnetic field are shown in Figure 1. The initial measurement of the fields in the center of the coil was initially between 0-30 Gauss. This may not create much difference because we could get similar results earlier with the plate reader [7] as well. All the magnetic arrangements were spread out in a big room to avoid the interference of magnetic field of the neighboring magnets.

The bacterial culture was originally obtained from American Type Culture Collection (ATCC 25923) and diluted 4 times to make the mother culture in nutrient broth. 0.1ml of mother culture was added to 20ml of nutrient broth for inoculation. All the optical density (OD) measurements were taken with 650nm wavelength using spec 20 on absorbance.

Different types of magnetic field effect on the growth of various bacterial species has previously been studied [8-11] in a constant and oscillating magnetic field. The damping field effect has been noticed for the first time. A relatively detailed investigation of the weak magnetic field effect on *S.aureus* is performed again to check the viability of use of magnetic field in medicine. For this purpose, we study the bacterial growth at room temperature keeping the constant concentration of nutrient broth and all other conditions like temperature and composition are maintained unchanged. We used the same nutrient broth which was prepared in a big flask and then distributed in different test tubes for experimental samples. All the study is done at the same time to avoid any possible environmental change.

## Results and Discussions

Figure 2 shows the growth pattern of up to 4 samples of bacteria which were all simultaneously grown under all of the similar conditions simultaneously. They all seem to follow the same growth curve. However, they may have a little different concentration after a long time because of the small changes in measurement time and the motion during measurement but the difference is still within the experimental error. The unique arrangement for continuous drainage of battery in *Figure 3 shows that the bacteria reached the static phase quickly and then even reached to death phase within 80 hours*. This was the only sample with survived magnetic field throughout. It reduced to 1 Gauss in the end. However, we could take measurements of optical density for the entire period for this sample and the strength of magnetic field has been reducing continuously throughout the entire period. It was also noticed that other three samples behaved similarly. However, the magnetic field vanished at different time for each sample and we could not keep track of time. In this case bacteria seems to acquire steady phase around 40 hours and seems to enter the death phase after 80 hours. Its maximum concentration is less than other cases of Figure 1 as well. Though the behavior of all four samples was similar for damping fields but results of other samples are not included due to uncertainty.

To compare the growth pattern, optical density and relative concentrations of bacteria in different magnetic fields, we plot averages of all the samples in Figure 4. It can be clearly seen that the growth of bacteria is enhanced in the constant electromagnetic field which is maintained to be constant throughout. Previous studied show a clear impact of magnetic field on the bacterial concentration of variable species [8-11]. However, the growth pattern is not affected with the magnetic field, regardless of the impact of the magnetic field on the change in concentration. However, it is easy to see in Figure 2 that damping field inhibits the bacterial growth.

Since the difference of concentrations in Figure 2 or 4 may be related to the change of initial concertation during the unstable growth period which leads to the difference of concentration and the death rate which may also be reduced due to the fast consumption of nutrients. Therefore, the preliminary study suggests that if the magnetic field is introduced for a short interval of time to reduce the concentration and then removed and then introduced again. It may assist to treat untreatable infections and may even completely treat it after some time without stronger chemical treatment.

It indicates that if the change of environment takes place, bacteria cannot grow during incubation period while its death rate may not be affected much and the concentration decreases. Now if the environmental change is made again to start another incubation period, growth will continue to reduce regularly. The successive environmental changes will lead to bacterial inhibition. This change can be managed by the short term exposure with the magnetic field or turning magnetic field on and off with intervals of several hours in an irregular fashion to trick bacteria. This type of process cannot be applied unless it is extensively tested on human infections because the complicated multifluid systems of human body may or may not support this method very well. But it is worth-proposing for further study on infected human body.

**Conclusion:**

The study of the effect of weak magnetic field on bacterial growth shows the initial inhibition period of a few hours and a measureable decrease in concentration of bacteria. If the magnetic field is turned on and off along with the change in nutrient concentration, it may lead to bacterial inhibition. A detailed study of the weak field use on human body may open up a new direction to treat antibiotic resistant infections.

**References**

1. Foster T. `Staphylococcus: Baron S, editor. Medical Microbiology. 4th edition. Galveston (TX): University of Texas Medical Branch at Galveston; 1996. Chapter 12. Available from: https://www.ncbi.nlm.nih.gov/books/NBK8448/
2. Easmon CSF, Adlam C: Staphylococci and staphylococcal infections. Vols 1 and 2. Academic Press, London, 1983.
3. Molly McAdow, Dominique M. Missiakas, Olaf Schneewind, `*Staphylococcus aureus* Secretes Coagulase and von Willebrand Factor Binding Protein to Modify the Coagulation Cascade and Establish Host Infections', J Innate Immun. 2012 Feb; 4(2): 141–148. Published online 2012 Jan 3. doi: 10.1159/000333447.
4. Lyon BR, Skurray R. Antimicrobial resistance in *Staphylococcus aureus*: genetic basis. Microbiol Reviews. 1987;51:88. [PMC free article] [PubMed]
5. Mohammad Abdul-Moqueet, Abdulla Alballawi and Samina Masood, `Study of Bacterial Response to Antibiotics in Low Magnetic Fields' BAPS.2017.MAR.K5.11, (2017).
6. Masood S (2017) Effect of weak magnetic field on bacterial growth. Biophys Rev Lett 12(04):177–186
7. S.Kang and **Samina Masood,** ` The Effect of Low Magnetic Fields on Bacterial Growth', (Presented in ASM meeting, May 2011). This work has been featured in Science News http://www.sciencenews.org/view/generic/id/74872/title/News_in_Brief_American_Society_for_Microbiology_meeting_
8. Ramon C, Ayaz M, Streeter DD (1981) Inhibition of growth rate of *Escherichia coli* induced by extremely low-frequency weak magnetic fields. Bioelectromagnetics 2(3):285–289
9. Gerencser VF, Barnothy MF, Barnothy JM (1962) Inhibition of bacterial growth by magnetic fields. Nature 196:539–541
10. Saleem I, Masood S, Smith D, Chu W-K (2018) Adhesion of gram-negative rod-shaped bacteria on 1D nano-ripple glass pattern in weak magnetic fields. MicrobiologyOpen. https://doi.org/10.1002/mbo3.640 (accepted for publication in MicrobiologyOpen)
11. Masood, S., Saleem, I., Smith, D. *et al.* Growth Pattern of Magnetic Field-Treated Bacteria. *Curr Microbiol* (2019) doi:10.1007/s00284-019-01820-7

# Figures

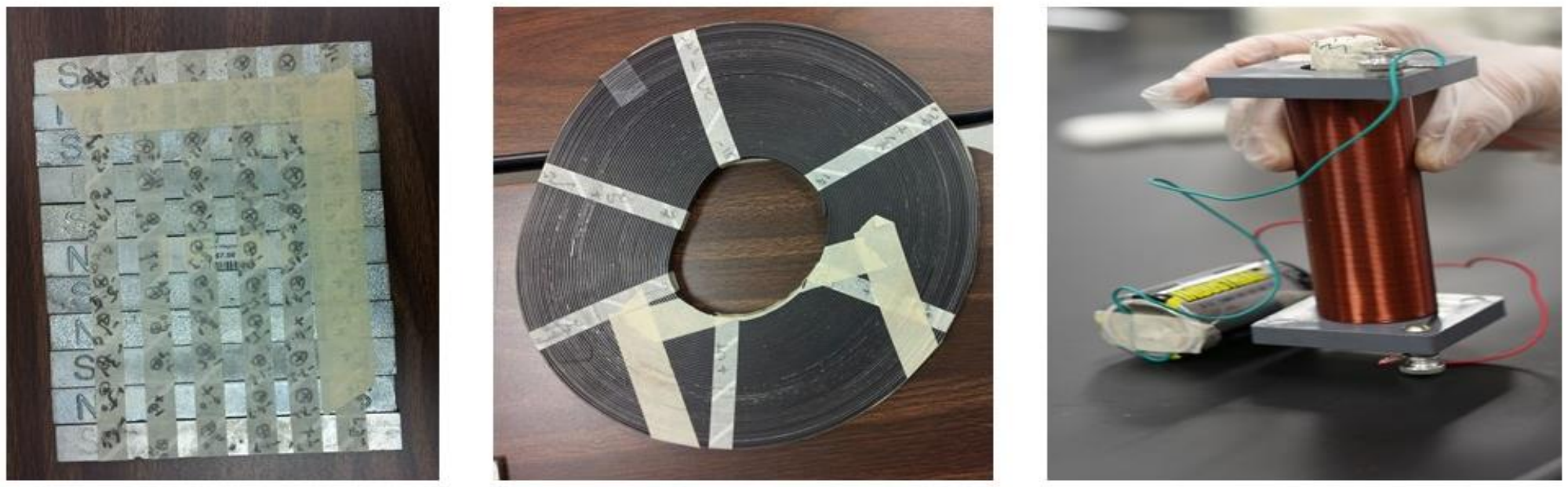

Fig 1. The magnetic field arrangements for the experiment. (a) static field arrangement, (b) Random field generating coil and (c ) Electromagnetic field inside the coil which perfectly fits a test tube attached with 3V battery.

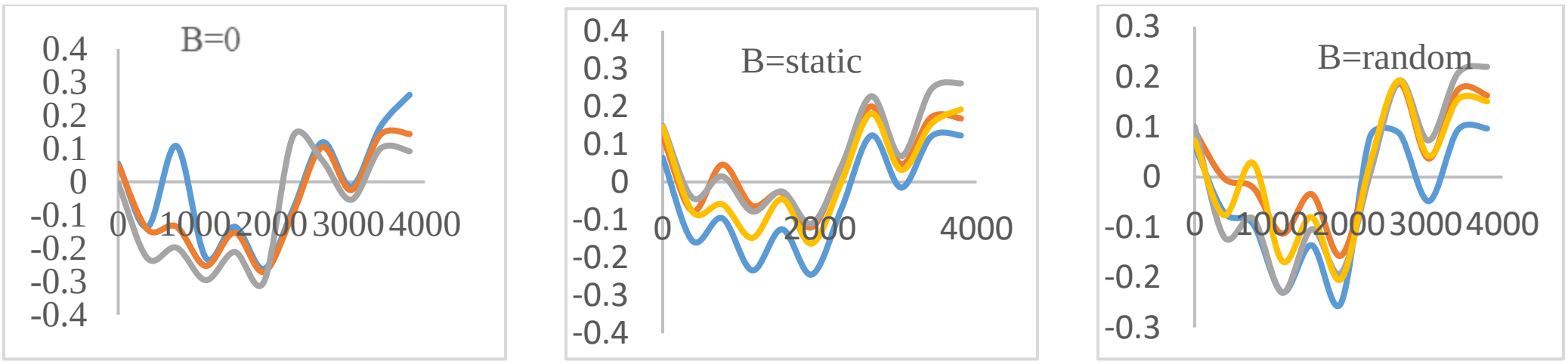

Fig 2. A growth pattern of *S.aureus* in various magnetic fields under particular conditions of magnetic field. Each case has four samples under same conditions. Optical density is plotted as a function of time in minutes.

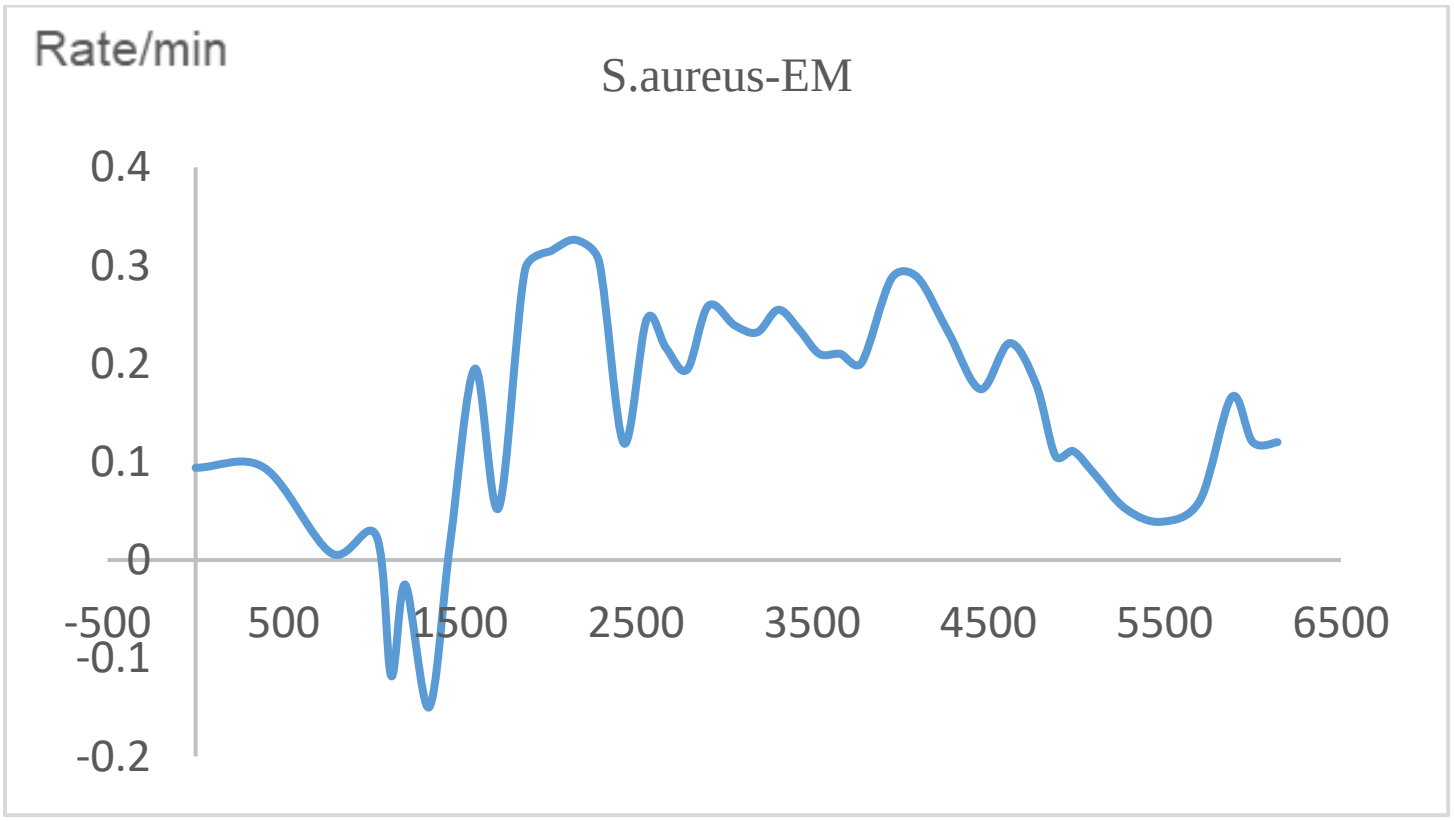

Fig 3. A special case of very weak magnetic field generated by a battery. Optical density is plotted as a function of time in minutes.

Figure 4

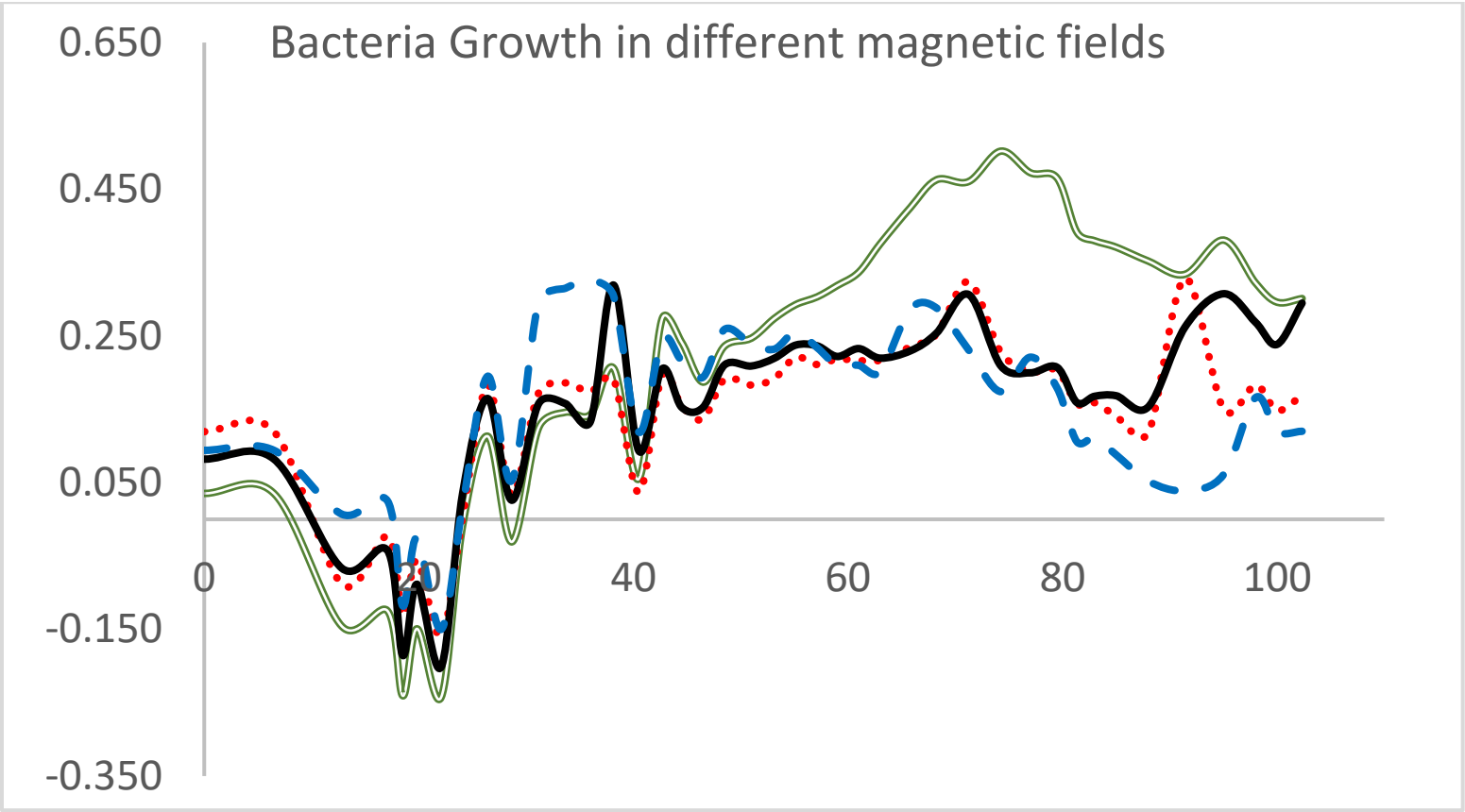

Fig.4. Comparison of growth rate of *S.aureus* with and without magnetic field. Each graph correspond to a different type of magnetic field with a comparable strength. Double orange line correspond to B=0,; Dotted red line indicates static field, solid black line plots bacteria grown over random magnetic field and dashed blue line B indicates damping electromagnetic field. Optical density is plotted as a function of time in minutes.